\documentstyle[12pt,aasms4]{article}
\input epsf
 
\def\k{km s$^{-1}$}
\def\pp{^{\prime\prime}}
\def\cm2{cm$^{-2}$}
\def\c3{cm$^{-3}$}

\begin{document}

\title{A molecular shell with star formation toward the supernova remnant 
G349.7+0.2 } 
 
\author{Estela M. Reynoso, \altaffilmark{1}} 
\affil{Instituto de Astronom\'\i a y F\'\i sica del Espacio,\\ CC 67  Suc 28,
(1428) Buenos Aires, Argentina}
\author{and Jeffrey G. Mangum}
\affil{National Radio Astronomy Observatory, 949 North Cherry Avenue, Tucson, 
AZ 85721-0655, USA}
 
\altaffiltext{1}{Member of the Carrera del Investigador Cient\'\i fico, 
CONICET, Argentina} 
\authoremail{ereynoso@iafe.uba.ar}
 
\begin{abstract}

A field of $\sim 38^\prime \times 38^\prime$ around the supernova remnant (SNR) 
G349.7+0.2 has been surveyed in the CO J=1--0 transition with the 12 Meter 
Telescope of the NRAO, using the On-The-Fly technique. The resolution of the 
observations is $54\pp$. We have found that this remnant is interacting with a 
small CO cloud which, in turn, is part of a much larger molecular complex, 
which we call the ``Large CO Shell''. The Large CO Shell has a diameter of 
about 100 pc, an H$_2$  mass of 9.3$\times 10^5$ M$_\odot$ and a density of 35 
\c3 . We investigate the origin of this structure and suggest that an old 
supernova explosion ocurred about $4 \times 10^6$ years ago, as a suitable 
hypothesis. 
Analyzing the interaction between G349.7+0.2 and the Large CO Shell, it is
possible to determine that the shock front currently driven into the molecular 
gas is a non-dissociative shock (C-type), in agreement with the presence of OH 
1720 MHz masers. The 
positional and kinematical coincidence among one of the CO clouds that 
constitute the Large CO Shell, an IRAS point-like source and an ultracompact H 
II region, indicate the presence of a recently formed star. We suggest that 
the formation of
this star was triggered during the expansion of the Large CO Shell, and suggest 
the possibility that the same expansion also created the progenitor star of 
G349.7+0.2. The Large CO Shell would then be one of the few observational 
examples of supernova-induced star formation.
\end{abstract}

 
\section {Introduction:}

The young, massive progenitor stars of Type II supernovae are born in large
molecular complexes. Since half of the Galactic supernova remnants (SNR) are
supposed to be of this type, it is expected that a similar number of SNRs be
in physical contact with their parent clouds. This contention is partially 
supported by CO observations (Huang \& Thaddeus 1986), which show a good 
positional correlation between large molecular complexes and SNRs. 
Thus, the interaction between SNRs and molecular clouds seems to be a 
common phenomenon in nature.

When a SN explodes near a molecular cloud, the expansion of the shock front 
into the cloud can accelerate relativistic particles, heat and compress the 
molecular gas, change its chemistry and produce turbulent mixing. Condensed 
clumps created by this mechanism may eventually end as new stars. However, 
very few cases provide direct evidence of SNR-molecular cloud interactions.
Morphological signatures are often invoked to prove a physical association
between an SNR and a molecular cloud (eg. Landecker et al. 1989, Reynoso et
al. 1995, Dubner et al. 1999), but the most convincing evidence for shocked
molecular gas are line broadenings (Frail \& Mitchell 1998, Reach \& Rho 1999)
or spectral wings (Seta et al. 1998). 

Another powerful tool indicative of SNR-molecular gas interaction is provided 
by the presence of OH 1720 MHz shock-excited masers. Within several OH 1720 MHz 
surveys (Frail et al. 1996, Green et al. 1997, Koralesky et al. 1998), in which 
around 150 Galactic SNRs have been observed, OH 1720 MHz masers have been 
detected toward nearly twenty of them.  In a recent paper, Reynoso \& Mangum 
(2001; hereafter Paper I) have surveyed the environs of three of these SNRs 
(G349.7+0.2, G16.7+0.1 and CTB 37A) in the CO J=1--0 transition, and found 
molecular clouds interacting with the shock fronts in all three cases. In this 
paper, we analyze the CO data in a large field around one of these remnants: 
G349.7+0.2. 
 
G349.7+0.2 is an incomplete clumpy shell, brighter to the south (Figure 1), with
a diameter of $2^\prime$. Based on HI observations (Caswell et al. 1975), this 
remnant is located beyond the tangent point, at a positive systemic velocity
greater that +6 \k , which places it also beyond the far side of the solar
circle. Clark \& Caswell (1976) suggest that the strong north-south
emission ridge observed in radio continuum is not a filled center but 
represents a non-uniform distribution of emission in a shell, which would be 
evidence for a strong interaction with the interstellar medium (ISM) at an 
early stage. In Paper I, it was shown that the OH 1720 MHz masers detected by 
Frail et al. (1996) toward this remnant, appear encircling a molecular cloud of 
$\sim 5.5$ pc in diameter, located at the same systemic velocity of the masers. 
The extension of the area covered by our survey allowed the detection of a 
much larger, ring-like CO feature, that includes the small molecular cloud 
interacting with G349.7+0.2. 

In this paper, we investigate the origin of this large structure and its 
connection with G349.7+0.2. In Section 2, we describe the observations. Section 
3 presents the main results, which are discussed in Section 4. In Section 5, we 
summarize our conclusions.

\medskip
\section {Observations and data reduction:}

Observations of the CO J=1--0 line (115.271204 GHz) were performed with the
NRAO\altaffilmark{2} \altaffiltext{2}{The National Radio Astronomy Observatory 
is a facility of the National Science Foundation operated under cooperative 
agreement by Associated Universities, Inc.} 12 Meter Telescope located on Kitt 
Peak, Arizona, on 1998 June 12-18 and 27-30. The beam size at this frequency is 
$54\pp$. The observational parameters are listed in Table {\ref{tbl1}}, 
including the noise of the final images. Observations were made using the 
On-The-Fly (OTF) technique, which 
allows efficient surveying of large areas. Three difrerent spectrometers were 
employed: two filter bank spectrometers with 128 channels and resolutions of 1 
MHz and 500 kHz, respectively, and a 768-channel autocorrelator with a 
resolution of 98 kHz. An absolute position-switching mode was used. The 
reference position was measured to be free of CO emission down to levels of 
less than about 0.5 K. Four maps, covering a square field of $\sim 38^\prime 
\times 38^\prime$, were obtained. 

Since data at the 12 Meter Telescope are calibrated by default to the T$^*_R$ 
scale, which is equivalent to the radiation temperature T$_R$ for a source much 
larger than the main diffraction beam of the telescope, it is often necessary 
to apply corrections which refer to sources of different sizes. For sources 
with sizes in the range $20\pp - 40\pp$, the efficiency factor which converts 
T$^*_R$ to T$_R$ is 0.85 based on historic measurements of the planets Jupiter 
and Saturn.

The AIPS package was employed for data processing. Baselines were subtracted
directly from the raw data with the task SDLSF. This task computes a first
degree polynomial using a unique range of line-free channels in the whole 
area. Because of the proximity of G349.7+0.2
to the Galactic plane, it was extremely difficult to find a suitable range of 
line-free channels. We thus decided to give priority to the baseline removal in 
the regions closest to the remnants and OH 1720 MHz masers, with possible 
detriment to the rest. Maps were averaged for each source using the task WTSUM, 
with $1/\sigma^2$ as weight, where $\sigma$ is the noise image. Due to the very 
high noise of one of the original maps, only three of them were kept in the 
average.

\placetable{tbl1}

\begin{deluxetable}{lccc}

\tablecaption{Observational Parameters\label{tbl1}}
\startdata
&\nl
\tableline
\tableline
Frequency [GHz]:&115.27 (2.6 mm)\nl
Number of channels:&768&128 &128\nl
Spectral resolution [kHz]&98 &500 &1000\nl
Spectral resolution [\k]&0.25 &2.6 &1.3\nl
Angular resolution [arcsec]:&53.6$\times$53.6\nl
Surveyed area [arcmin]:& $\sim 38 \times 38$\nl
Observing mode :& absolute position switched\nl
&On-The-Fly (OTF)\nl
Central position [RA (2000)]& 17 17 59.90 \nl
{\hskip 3 cm} [decl. (2000)]:& -37 26 9.0\nl
Reference position [RA (2000)]& 17 03 59.23\nl
{\hskip 3 cm} [decl. (2000)]:& -37 27 8.7\nl
Number of maps:&3\nl
Observing time [hours]:&4.1\nl
Central velocity (LSR)  [\k ]:&--40\nl
rms [K]:&0.6&0.4&0.4\nl
\enddata

\end{deluxetable}

\medskip
\section {Results:}

Figure 2 displays the CO distribution in the range 12.6 to 21.2 \k . At $v\sim 
+14.5$ \k, there appears a chain of clouds forming an arc-like feature opened 
to the northeast. At $v\sim +16.5$ \k, this arc merges with the bulk of the CO 
emission at the northeast, outlining a circle that fades away at $v\sim +20.5$ 
\k. The cloud interacting with G349.7+0.2 is part of this molecular complex.
Figure 3 shows the CO emission integrated over the velocity interval +14.5 to
+20.5 \k . The extended CO structure is remarked by a 700 $\times$ 770 arcsec 
ellipse that fits the brightest peaks. In what follows, we will refer to the 
large CO structure as the ``Large CO Shell'', and will identify individual 
smaller clouds within this structure with numbers. 

One of the individual clouds, hereafter ``cloud 1'', has been reported in
Paper I. Cloud 1 has a systemic velocity of +16.5 \k , and is positionally 
coincident with G349.7+0.2. Based on the presence of OH 1720 MHz 
shock-excited masers at the same position and velocity, as well as infrared 
emission from shock-heated dust arising from the same location, it was
concluded that the SNR is interacting with this molecular cloud.

As in Paper I, masses are estimated by integrating $N_{H_2}$ over the solid 
angle subtended by the CO emission feature. To calculate the H$_2$ column 
density, the relationship $X=N(H_2)/W_{CO}$ is used, where $W_{CO}$ is the 
integrated CO line intensity, $W_{CO}=\int{T_R(CO)}\, dv$, in K \k , and where 
we have divided our antenna-based T$^*_R$ temperatures by the efficiency factor 
of 0.85 to place them on the T$_R$ scale appropriate for sources with sizes in 
the range $20\pp-40\pp$. 

Using a mean molecular weight per H$_2$ molecule 
of 2.72 m$_H$ (Allen 1973), the mass of a cloud can be expressed as:
\begin{equation}M=5.7\times 10^{-21}\, X\, W_{CO}\, \theta^2\, D^2\ \ M_
\odot ,\end{equation}
\noindent where $D$ is the distance in kpc, and $\theta$ is the angular radius 
of the cloud in arcmin, computed as half the measured 
diameter deconvolved by the beam size. Assuming spherical volumes, the density
can be estimated as:
\begin{equation}n_{H_2}={{4.1\times 10^{-19}\, X\, W_{CO}}\over{D\, 
\theta}}\ \ {\rm cm}^{-3}.\end{equation}

Based on the close agreement between the systemic velocities of cloud 1 and the 
Large CO Shell, the distance adopted for the whole molecular complex will be 
the kinematical distance to cloud 1, $\sim 23$ kpc, which corresponds to a 
Galactocentric distance of $R=14$ kpc (Paper I). Therefore, the ratio $X$ is 
assumed to be $8\times 10^{20}$ \cm2 (K \k )$^{-1}$ (Digel, Bally \& Thaddeus 
1990). The derived parameters may be over estimated by a factor of 2 (see Paper 
I). The mass and H$_2$ density of the Large CO Shell can be roughly estimated 
to be about 9.3$\times 10^5$ M$_\odot$ and 35 \c3, respectively. In this 
computation, an outer diameter of $\sim $ 870 arcsec and an inner diameter of 
$\sim $ 500 arcsec were assumed. At the distance assumed, the outer diameter of 
the Large CO Shell is $\sim$ 100 pc. Although the density is somewhat low, both 
mass and size place this structure in the category of giant molecular clouds 
(Goldsmith 1987). 

With the aim of exploring if star formation is taking place in the Large CO
Shell, we have searched for protostellar candidates in the IRAS Point Source 
Catalogue that fulfill the following selection criteria: (1) $S_{100}\geq 20$ 
Jy, (2) $1.2 \leq S_{100}/S_{60}\leq 6.0$, and (3) $S_{60}/S_{25}\geq 1$, where 
$S_\lambda$ denotes the (uncorrected) IR-flux density in the wavelength $\lambda \, 
\mu$m. The first criterion selects only strong sources, while the second and 
third discriminate against cool stars, planetary nebula and cirrus clumps 
(Junkes, F\"urst \& Reich 1992). However, these selection criteria may include 
not only protostars but also dust heated in SNR shocks (Arendt 1989). Three 
IRAS sources are found in the field covered by the present survey (white stars 
in Fig. 3), two of them coincident with CO clumps: IRAS 17146-3723 at cloud 1, 
and IRAS 17147-3725 at a CO concentration centered approximately at RA=$17^h 
18^m 10^s.0$, decl=$-37^\circ 28^\prime 25\pp$ (J2000), hereinafter named 
``cloud 2''. The superposition of these two IR sources with CO maxima is very 
unlikely to be by chance, and their connection with the CO emission will be
discussed in the next Section. 

In Table  {\ref{tbl2}} we summarize the parameters estimated for each CO
structure. The parameters of cloud 1 have been also reported in Paper I. The 
second column quotes the approximate central position of the structures in 
equatorial coordinates referred to J2000. The third column lists the radii $r$ 
after deconvolution with the beam size. For the Large CO Shell, only the outer 
radius is given. The fourth and fifth columns contain the systemic velocity 
$v_{sys}$ and the FWHM velocity $\Delta v$, both in \k. The computed mass and 
H$_2$ densities are given in the last two columns. 

\placetable{tbl2}

\begin{deluxetable}{lcccccc}

\tablecaption{Observed and derived parameters for CO structures 
associated with G349.7+0.2\label{tbl2}}
\tablehead{
        \colhead{structure} &
        \colhead{central coordinates} &
        \colhead{r } &
        \colhead{$v_{sys}$}&
	\colhead{$\Delta v$}&
	\colhead{M} &
	\colhead{$n_{H_2}$}\nl
        \colhead{} &
        \colhead{RA, decl. (J2000)} &
        \colhead{(arcsec)} &
        \colhead{(\k)}&
	\colhead{(\k)}&
	\colhead{(M$_\odot$)} &
	\colhead{(\c3)}}
\startdata
Cloud 1&17 18 0.0, -37 26 36&24&+16.5&4.0&1.2$\times 10^4$&1080\nl
Cloud 2&17 18 10.0, -37 28 25&33&+17.3&4.5&2.2$\times 10^4$&790\nl
Large CO Shell&17 18 22.3,  -37 22 55&435&+17.5&6.3&9.3$\times 10^5$&35\nl
\enddata

\end{deluxetable}

\section {Discussion:}

\subsection{The Large CO Shell and the SNR G349.7+0.2}

There is compelling evidence that G349.7+0.2 is interacting with a
molecular cloud. Moreover, our present results suggest that the SN event 
occurred inside the Large CO Shell, thus confirming the statement of Clark \& 
Caswell (1976) that the expansion has been influenced by a dense environment. 
Nevertheless, it is unlikely that cloud 1 has interacted with the 
shock front in early stages, since stars are more probable to explode in 
the interclump gas than in dense clumps (Chevalier 1999). 

The radio structure of G349.7+0.2 can be fitted by two overlapping rings, with 
one smaller and brighter than the other. This characteristic similar to many
other SNRs, can be easily explained if the smaller ring corresponds to the
side of the expanding shell encountering the higher density (Manchester 1987). 
We therefore propose a scenario in which the SN shock front has expanded into
a density gradient; the smaller half of the shell has expanded into higher 
density gas and is now running through cloud 1. 

Assuming a distance of 23 kpc to G349.7+0.2, the radii of the large and small
rings are $R_l \sim 9$ pc and $R_s \sim 6$ pc respectively. Based on Fig. 3,
the small ring appears to be expanding into the Large CO Shell, while the large 
ring is expanding into a region of low density. If the energy was conserved 
during expansion, then $n R^2/t^2=const.$ (e.g. Dubner et al. 1992), and we can 
expect that the distances $R_l$ and $R_s$ travelled by the shock front during 
the lifetime of the remnant be related by $(R_l/R_s)^2=n_s/n_l$, where $n$ 
refers to the ambient medium into which each of the two rings is expanding. 
Replacing $n_s$ by 35 \c3 \ (Table {\ref{tbl2}}), we deduce that $n_l=15$ \c3.

In an inhomogeneus medium, an SNR can be undergoing different evolutive stages
in different places at the same time. Here, the expansion of the larger ring is
analyzed. Cioffi, McKee \& Bertschinger (1988) determined the radius of an SNR
at the onset of the radiative phase as:
\begin{equation}R_c=14 \left( {E_{51}^{2/7}\over {n_0^{3/7}\, \xi_m^{1/7}}}
\right),\end{equation}
where $\xi_m\simeq 1$ for solar abundances. Adopting a canonical value of 
$E_0=10^{51}$ ergs, and assuming an ambient density of 15 \c3, equation (3)
yields $R_c=12$ pc. In spite of the uncertainties involved in this 
determination, it can be considered then that the radiative phase has not been
achieved yet, and therefore the use of the Sedov equations is appropriate.

The equations that govern the expansion of an SNR in the Sedov phase are
extracted from Cox (1972): 
\begin{equation}R_s=13.6\, t_4^{2/5}\,\left({E_{51}\over n_0}\right)^{1/5}\ 
{\rm pc}\end{equation}
\begin{equation}V_s=535\, t_4^{-3/5}\,\left({E_{51}\over n_0}\right)^{1/5}\
{\rm km\ s^{-1} } \end{equation}
From equation (4), the age of the remnant is 1.4$\times 10^4$ yrs, and thus
the shock velocity is $\sim 260$ \k. Using once again the relationship
$(V_l/V_s)^2=n_s/n_l$, we find that the velocity of the shock front driven
into the higher density gas of the Large CO Shell is $V_s=170$ \k \ which,
when entering cloud 1, can be slowed down to 33 \k. At velocities lower than
$\sim 50$ \k, shocks are of C-type (Neufeld \& Dalgarno 1989), thus this 
estimate is consistent with the theoretical prediction by Lockett, Gauthier \&
Elitzur (1999) that the conditions under which 1720 MHz OH maser emission is 
produced can exist only if shocks are of C-type.

\medskip
\subsection{Star formation in the Large CO Shell}

Based on the morphology, we suggest that the Large CO Shell is the remains of 
an expansive event. In what follows, we try to explain the origin of this
structure. The hypothesis of a wind-driven bubble can be discarded since 
a literature search revealed no early-type stars or associations within the
area subtended by the Large CO Shell. Therefore, as an alternative, we will 
investigate the hypothesis that the expansive event was a SN explosion. 
Assuming that the mass of the Large CO Shell was initially distributed
uniformly in a sphere of radius $435\pp$ ($\sim 50$ pc), then the original 
density was $n_0=30$ \c3. Since the Large CO Shell extends over a velocity 
interval of about 6 \k, we adopt a velocity expansion of 3 \k \ as a working 
hypothesis.  Taking into account that in the radiative phase, which is the last 
stage in the evolution of an SNR before dissipating in the ISM, the expansion 
law is $R\propto t^{1/4}$, an upper limit to the age of the shell can be 
obtained as $t=R/4\,V\simeq 4\times 10^6$ yrs. This age is sufficient for the 
formation of protostars. In the next paragraphs, the origin of the two IRAS 
sources found in coincidence with CO peaks will be discussed.

Because of its spectral characteristics, IRAS 17147-3725 is catalogued as an 
ultra-compact (UC) H II region (Wood \& Churchwell 1989).  In Fig. 3, it can be
seen that IRAS 17147-3725 is positionally coincident with a weak, unresolved
radio continuum point source. This continuum source has a peak flux more 
than 20 times above $\sigma$, and has an integrated flux of $S_\nu=0.025$ Jy at 
1.41 GHz. We propose that this source is the radio continuum counterpart of the 
UC-H II region IRAS 17147-3725.  

Compact H II regions are created when ultraviolet emission from newly formed 
stars ionize the neighboring gas. Also, the surrounding dust is heated and 
reradiates the energy at infrared wavelengths. Therefore, the IRAS source 
17147-3725 is very likely dust heated by a recently formed star. To trace the 
high density cores where high mass stars are born, Bronfman et al. (1996) 
observed UC-H II regions in the CS J=2-1 line, and found that IRAS 17147-3725 
has a peak at +17.6 \k , which is in coincidence with the systemic velocities 
of cloud 2 and the Large CO Shell (Table {\ref{tbl2}}).

The Lyman continuum (LyC) flux from the ionizing star in the UC-H II region can 
be derived from the radio continuum flux as:
\begin{equation}N_{LyC}\simeq 7\times 10^{46}\, \nu^{0.1}\, S_\nu\, D^2,\end{equation}
where $\nu$ is given in GHz, $S_\nu$ in Jy and $D$ is the distance in kpc
(Mezger, Smith \& Churchwell 1974). A temperature of $T_e=10^4$ K and a ratio
$^4$He/H=0.1 have been assumed. The flux obtained, $N_{LyC}\simeq 9.6\times
10^{47}$ photons s$^{-1}$, is consistent with an O9 - O9.5 ZAMS type star
(Panagia 1973). 
On the other hand, following Helou, Soifer \& Rowan-Robinson (1985), the 
luminosity of the IRAS point source can be estimated from the fluxes at 60
$\mu$m and 100 $\mu$m according to:
\begin{equation}L_{FIR}=4\pi D^2\, FIR=0.385 D^2 (2.58\, S_{60}+S_{100})\,
L_\odot=4.1\times 10^5\, L_\odot,\end{equation}
where $FIR$ is the integrated flux from the band between 42 $\mu$m and 122 
$\mu$m. We can compute, then, the parameter $\Theta=L(IR)/(N_{LyC}\, E_{LyC}\,
\alpha)$ that characterizes a compact H II region (Panagia 1974), where 
$E_{LyC}$ is the energy of a single Lyman-$\alpha$ photon and $\alpha$ is
the ratio of the total luminosity of the star to the expected Lyman-$\alpha$
luminosity. For a O9.5 type star, $\alpha=13.2$ (Panagia 1973), thus
$\Theta\simeq 8$. High values of $\Theta$ ($\Theta>1$) indicate that the dust
is still mixed with the ionized gas to an appreciable extent. The value 
obtained here implies therefore that this UC-H II region is a very young object.

Let us now discuss the origin of IRAS 17146-3723, the source coincident with
cloud 1. Condensed molecular matter can form a protostar after the matter has 
cooled, which occurs in a period of the order of $10^6$ years. Such a period is 
too long compared to the age of G349.7+0.2 deduced in Section 4.1, thus 
this IRAS source cannot be a protostar triggered by G349.7+0.2. We suggest,
therefore, that the IR emission from this source arises from dust heated by
the shock front of G349.7+0.2.  The OH 1720 MHz masers detected by Frail et 
al. (1996) towards this remnant provides additional support to the presence of 
shocked interstellar gas. 

The Large CO Shell shares some common characteristics with another SNR, 
G54.4-0.3, 
which is claimed to be a case of SN induced star formation: both present a CO 
shell slowly expanding at $\sim 4$ \k, in an ambient medium with density $n_0=
30$ \c3 , and both have an IRAS source coincident with a compact H II region 
created by an O9 type star (Junkes et al. 1992). The main difference is that 
the CO shell of G54.4-0.3 coincides with a radio continuum shell, while 
available radio surveys show no continuum emission at the position of the Large 
CO Shell. According to the $\Sigma$-$D$ relation of Case \& Bhattacharya 
(1998), the surface brightness that should be observed is $\Sigma\simeq 4\times 
10^{-22}$ W m$^{-2}$ Hz$^{-1}$ sr$^{-1}$. Such a low surface brightness is 
probably missed by Galactic radio surveys (see Green 1991). Nevertheless, both 
the age derived above and the low expansion velocity, suggest that the shock 
has disappeared and the Large CO Shell is an SNR in the phase of dissipation. 
Besides, due to the old age of this remnant, there is no need for an 
alternative explanation for the origin of the star that created the UC-H II 
region IRAS 17147-3725. In the case of G54.4-0.3, instead, Junkes et al. (1992) 
need to include a stellar wind previous to the SN explosion to explain how the 
protostar detected as an IRAS point source was triggered, since the SNR was not 
old enough.

In summary, the scenario proposed here is that a SN exploded some $\sim 4\times
10^6$ years ago at approximately RA=$17^h 18^m 22^s.3$, decl=$-37^\circ 
22^\prime 55\pp$ (J2000), in a molecular cloud with density $n_0=30$ \c3 . The 
SN shock wave triggered the formation of an O9 - O9.5 type star, currently 
detected as the IR source IRAS 17147-3725. It is possible that the progenitor
star of G349.7+0.2 has also been triggered by the expansion of the Large CO 
Shell. The Large CO Shell, therefore, constitutes a new candidate for 
SN induced star formation. Even though the evidence is not conclusive, it is 
important to note that the case most strongly claimed to be SN induced star 
formation, is based on the spatial coincidence between the association CMa R1 
and an optical ring which, together with an HI expanding shell, is interpreted 
as an old SNR (Herbst \& Assousa 1977). However, the hypothetical SNR could 
never be confirmed. 

\section {Conclusions:}

The surroundings of the SNR G349.7+0.2 have been investigated in the CO J=1-0 
transition. Our observations allowed us to determine the density distribution 
of the ISM into which this remnant is expanding. A large, ring-like structure
has been detected, which we called ``the Large CO Shell''. It was found that
``cloud 1'', the CO cloud interacting with G349.7+0.2 (Paper I), is part of the
Large CO Shell. We suggest that this molecular shell is the remnant of an old
supernova explosion, now in the phase of dissipation.

We analyzed the interaction between G349.7+0.2 and the Large CO Shell, and
concluded that the age of the former is about 1.4$\times 10^4$ years. The
shock driven by G349.7+0.2 into cloud 1 was confirmed to be of C-type, in 
agreement with the conditions requiered for OH 1720 MHz masers to exist  
(Lockett et al. 1999).

Three IRAS point like sources with spectral characteristics typical of 
protostars were found in the Large CO Shell. One of them, IRAS 17146-3727, 
is a clump of dust heated by the shock front (Paper I). Another one, IRAS 
17147-3725, is located at the position of a CO peak, and coincides with an 
UC-H II region, probably produced by an O9-9.5 type star.

In summary, we propose that the explosion of a star, about $4\times 10^6$ years 
ago, in a medium with density 30 \c3, pushed the surrounding gas forming what 
we see now as the Large CO Shell and triggered, during its expansion, the birth 
of an O9 - O9.5 type star detected as the IR source IRAS 17147-3725. It is
possible that the formation of the progenitor star of G349.7+0.2 was also 
induced in the same way. High sensitivity radio-continuum observations are 
needed to investigate whether the Large CO Shell has a counterpart in 
synchrotron emission and to measure the strength and orientation of magnetic 
fields in this direction. This information is crucial in determining a SN 
origin for this structure.

\bigskip

We are grateful to the NRAO for allocating time in the 12  Meter Telescope
for this project. We also thank C. Salter for providing us his continuum image 
of G349.7+0.2.  E.M.R. acknowledges travel grant 1526/98 from CONICET 
(Argentina) for visiting the 12 Meter Telecope, as well as NRAO support and 
staff assistance during her visit. This research was partially funded through 
a Cooperative Science Program between CONICET and NSF and through CONICET grant 
4203/96.

\section{Figure captions}

\figcaption[Figure1.ps]{A radio continuum image of G349.7+0.2 at 1.4 GHz
obtained with the VLA, kindly provided by C. Salter. The grayscale is indicated 
in mJy/beam on top of the image, while the contours are expressed as a 
percentage of the peak flux density (0.48 Jy/beam) at 1, 2, 3, 5, 10, 20, 30, 
40, 50, 60, 70, 80, 90, and 99\%. For better contrast, white contours are used 
over dark shadows. White crosses indicate the position of the OH 1720 MHz
masers detected by Frail et al. (1996). The beam, indicated in the bottom left 
corner, is 19$\pp.4 \times 4\pp.5$, P.A.=$-22^\circ$.}

\figcaption[Figure2.ps]{CO emission between 12.6 and 21.2 \k. The greyscale is
in Kelvins, and is shown on top of the images. The beam size, 54$\pp \times
54\pp$, is shown in the bottom right corner of the first frame. The velocity
of each frame is indicated in the top right corner. All velocities are referred
to the LSR.}

\figcaption[Figure3.ps]{CO emission  integrated between +14.5 and +20.5 \k . 
The greyscale is in K \k , and is shown on top of the image. The original image 
was convolved to a 60$\pp \times 60\pp$ beam, plotted in the bottom right 
corner. An ellipse is fitted to the CO peaks. The white stars indicate the 
position of three IRAS point sources, discussed in the text. A few 
representative contours of the continuum emission from G349.7+0.2 are included 
as grey solid lines. The 3$\sigma$ noise level is 6.5 K \k.}

\end{document}